\documentclass{article}
\usepackage{spconf,amsmath,graphicx,booktabs,enumitem}
\usepackage{xcolor}

\title{Unsupervised Fine-Tuning Data Selection for ASR Using Self-Supervised Speech Models}
\name{Reem Gody, David Harwath}
\address{The University of Texas at Austin \\ \textit{reemgody@utexas.edu, harwath@utexas.edu}}
\begin{document}
\ninept
\maketitle
\begin{abstract}

Self-supervised learning (SSL) has been able to leverage unlabeled data to boost the performance of automatic speech recognition (ASR) models when we have access to only a small amount of transcribed speech data. However, this raises the question of which subset of the available unlabeled data should be selected for transcription.
Our work investigates different unsupervised data selection techniques for fine-tuning the HuBERT model under a limited transcription budget.
We investigate the impact of speaker diversity, gender bias, and topic diversity on the downstream ASR performance. We also devise two novel techniques for unsupervised data selection: pre-training loss based data selection and the perplexity of byte pair encoded clustered units (PBPE) and we show how these techniques compare to pure random data selection. Finally, we analyze the correlations between the inherent characteristics of the selected fine-tuning subsets as well as how these characteristics correlate with the resultant word error rate. We demonstrate the importance of token diversity, speaker diversity, and topic diversity in achieving the best performance in terms of WER. 
\end{abstract}
\section{Introduction}
\label{sec:intro}

Self-supervised speech models like HuBERT and wav2vec 2.0~\cite{baevski2020wav2vec,hsu2021hubert} have achieved very low WER when pre-trained on a large dataset of untranscribed speech and fine-tuned on as little as 1 hour of transcribed data. This motivates using these models for automatic speech recognition for low-resource scenarios where we may have access to a moderate to large amount of untranscribed speech but a finite budget available for data transcription. However, this raises the question of how to optimally choose which subset of the data should be transcribed for fine-tuning the model. Using small amounts of data in fine-tuning is associated with the risk of having high variance in the WER at test time, depending on the characteristics of the subset selected for fine-tuning. Our goal in this paper is to investigate different selection criteria for choosing the fine-tuning subset, and their effect on downstream ASR performance. In our setup, we assume that we have a large pool of unlabeled in-domain data and a limited transcription budget, e.g. 10 hours. We need to select a subset of this data pool to transcribe and use for fine-tuning the model. We probe the impact of speaker diversity, gender bias and topic diversity on the model performance. Moreover, we devise two novel techniques for unsupervised data selection: Pre-trained loss based data selection and Perplexity of byte pair encoded clustered units (PBPE) and we show how these techniques compare to pure random data selection. Section \ref{sec:relatedwork} discusses some related work. Section \ref{sec:proposedcriteria} explains our proposed criteria for data selection. Section \ref{sec:Experiments} describes our experimental setup, the results we achieved, and provides some analysis. Finally, section \ref{sec:conclusions} summarizes our conclusions and future work.

\section{Related work}
\label{sec:relatedwork}

Training data selection plays an important role in efficient training, dealing with a limited transcription budget, and domain adaptation. 
Unsupervised techniques have been considered for selecting data for both supervised or semi-supervised learning. The work in~\cite{park2022unsupervised} uses contrastive loss ratios between two models trained on general and target data to select an optimal subset for transcription. They aim to decrease supervised training time by selecting a portion of the available data without suffering performance degradation. The authors in 
\cite{afshan2021sequence} propose using confidence scores to select an optimal adaptation set, in order to minimize the transcription budget for supervised adaptation and choose the best hypotheses for semi-supervised adaptation, while maximizing ASR gains.
In the self-supervised paradigm, downstream performance on a target domain can be optimized by the optimal selection of the pre-training data or the fine-tuning data. \cite{lu2022unsupervised} use a contrastive data selection method applied to the learned discrete tokens for selecting data for pre-training an ASR model. They show significant improvement when selecting pre-training data that is matched to the target domain compared to pre-training using a full dataset of 1 million youtube hours.  \cite{kawakami2020learning} and \cite{hsu2021robust} analyze the effect of domain-shift between the pre-training and testing phases. Our work looks at a different angle of the problem. Given a pool of unlabeled data from a given target domain that is used for pre-training a model, what are the different ways by which we can select a subset of this data to transcribe and then use to fine-tune the model?  We assume a fixed transcription budget and analyze how the performance varies as we change the selection criteria.

\section{Proposed Selection Criteria}
We try several criteria for data selection using the Librispeech~\cite{panayotov2015librispeech} dataset with a limited transcription budget of 10 hours of speech and compare these criteria to pure random selection. We investigate the effect of speaker diversity by forcing a specified number of speakers into the fine-tuning subset. Moreover, we probe the impact of gender bias by selecting subsets with either female or male speakers only. We examine the impact of topic diversity by limiting our selection to a specified number of audiobooks. Moreover, we test the effect of batching short utterances vs long utterances on the downstream performance. We investigate two novel techniques for unsupervised in-domain data selection: pre-training loss based data selection (PL-based) and perplexity of byte pair encoded clustered units (PBPE). We enumerate all of our data selection criteria in Table~\ref{tab:selection_criteria}.

\label{sec:proposedcriteria}
\subsection{Pre-training loss based data selection}
We base our experiments on the HuBERT model and investigate the use of the HuBERT pre-training loss function as a means of data selection. Similar to wav2vec 2.0, HuBERT selects \textit{p}\% of the time steps as start indices for masking, and then spans of \textit{l} time steps are masked. Using targets derived via K-means clustering of MFCCs or features extracted using a previous snapshot of the model, the cross-entropy loss is then computed over the masked and the unmasked time steps as \begin{math}L_m\end{math} and \begin{math}L_u\end{math} and the weighted sum of both is taken as the final loss as defined in the following equation:
\begin{equation}
 L = {\alpha L_m+(1-\alpha) L_u}
\end{equation}
For our data selection, we compute the score of each utterance as a function of the pre-training loss. First, we compute the utterance score as the average pre-training loss over all the masked frames in the utterance. We sort the utterances ascendingly based on the computed scores and select the first 10 hours or the last 10 hours. Since the masking is inherently random, we hypothesize this may lead to a criterion that is close to pure random selection as the same utterance will get some different score each time we compute the loss. In light of this, we also consider turning off the mask and computing the utterance score as the average cross-entropy loss over all the unmasked frames. The advantage of this second method is that we get a deterministic score for each utterance. We sort the utterances ascendingly based on the computed scores and randomly sample utterances from both the top and the bottom 15\% of the data. 


\begin{table*}[h!]
    \centering
    \footnotesize
    \begin{tabular}{|l|l|}
        \toprule Criterion & Description \\
        \hline
         PUR\_RND & Sample 10 hours randomly \\
         \hline
         GNDR\_DIV& Sampled subset contains 24 speakers of one gender ( male / female )\\
         \hline
         UTTLN\_DIV\_RND\_LNG\_DUR\_TAIL & Sample from the 15\% of the utterances with the longest duration\\
         \hline
         UTTLN\_DIV\_RND\_SHRT\_DUR\_TAIL & Sample from the 15\% of the utterances with the shortest duration\\
         \hline
         UTTLN\_DIV\_RND\_MIDDLE\_DUR & Sample from the middle 15\% of the utterances in terms of duration\\
         \hline
         SPK\_DIV\_RND & Sample utterances for a specified number of speakers (24 - 96)\\
         \hline
         BK\_DIV\_RND & Sample utterances from a specified number of books (16 - 64) \\\midrule
         PRETRAIN\_U\_LOSS\_AVG\_NO\_MASK & 
         Compute the average unmasked pre-training loss for each utterance after turning off the mask.\\
            & Sample from the 15\% with the lowest loss (HEAD) vs the 15\% with the highest loss (TAIL).\\
         \hline
         
         PRETRAIN\_M\_LOSS\_AVG & 
        Compute the average masked pre-training loss for each utterance.\\
  & Sample from the 15\% with the lowest loss (HEAD) vs the 15\% with the highest loss (TAIL).\\
         \hline
         PERPLEXITY\_5k\_LM\_15 & 
            Use PBPE to compute utterance score.\\
            & Sample from the 15\% with the lowest score (HEAD) vs the 15\% with the highest score (TAIL).\\
         \hline
         PERPLEXITY\_5k\_LM\_40\_MIDDLE & Sample utterances with PBPE from the middle 40\% of the data \\\bottomrule
         
    \end{tabular}
    \caption{Description of data selection criteria for 10 hour fine-tuning subsets}
    \label{tab:selection_criteria}
\end{table*}

\subsection{Perplexity of Byte Pair Encoded clustered units (PBPE)}
\label{pbpe}
HuBERT uses K-means clustering to generate noisy labels for the pre-training step. The primary reason this approach works is that even though the labels are noisy, they tend to be consistent~\cite{hsu2021hubert}. We hypothesize that these labels are highly correlated to the tokens in the actual text transcripts in a way that enables us to utilize the same algorithms used for labeled data selection. 
First, we apply run-length encoding to collapse consecutive repetitions of the frame-level HuBERT cluster labels. 
Next, we use sentencepiece \cite{kudo2018sentencepiece} to train a BPE model on the run-length encoded sequences with a vocab size of 5k. Finally, we tokenize each sequence using that BPE model. We use fairseq \cite{ott2019fairseq} to train a 1-Layer LSTM language model with 512 hidden units over these unit-BPE sequences. We use the language model to compute the perplexity of each utterance in Librispeech train set. We then sort the utterances based on their perplexity and sample utterances from the top 15\% and the bottom 15\% of the whole train set to compare both criteria. 

 


\section{Experimental Setup and results}
\label{sec:Experiments}

\subsection{Model and Data}
In our experiments, we use the HuBERT base model which is pre-trained on the full 960 hours of Librispeech \cite{panayotov2015librispeech} and is available on fairseq. Moreover, we use the same K-means clustering model with 500 clusters that is trained on the latent representations of HuBERT's 9th transformer layer after pre-training for 2 iterations.

For our data selection experiments, we use the full 960 hours of Librispeech as our data selection pool. Because each of our data selection criterion still utilize random sampling in some form, we prepare 8 fine-tuning subsets,of 10 hours each. We experiment with different selection criteria for these subsets to probe how the model would behave under a differing number of speakers, differing number of topics (books), and speaker gender bias. We also examine the impact of grouping utterances with similar lengths to see whether this would help the model learn better given the same number of updates and same maximum number of tokens per batch. In addition to this, we experiment with our proposed criteria: pre-training loss based data selection and PBPE.
Table \ref{tab:selection_criteria} summarizes the different data selection criteria that we used in our fine-tuning experiments. We use Librispeech dev-other for validation and we test our models on both Librispeech test-clean and test-other.

\begin{figure*}[t!]
    \centering
    \includegraphics[width=\linewidth]{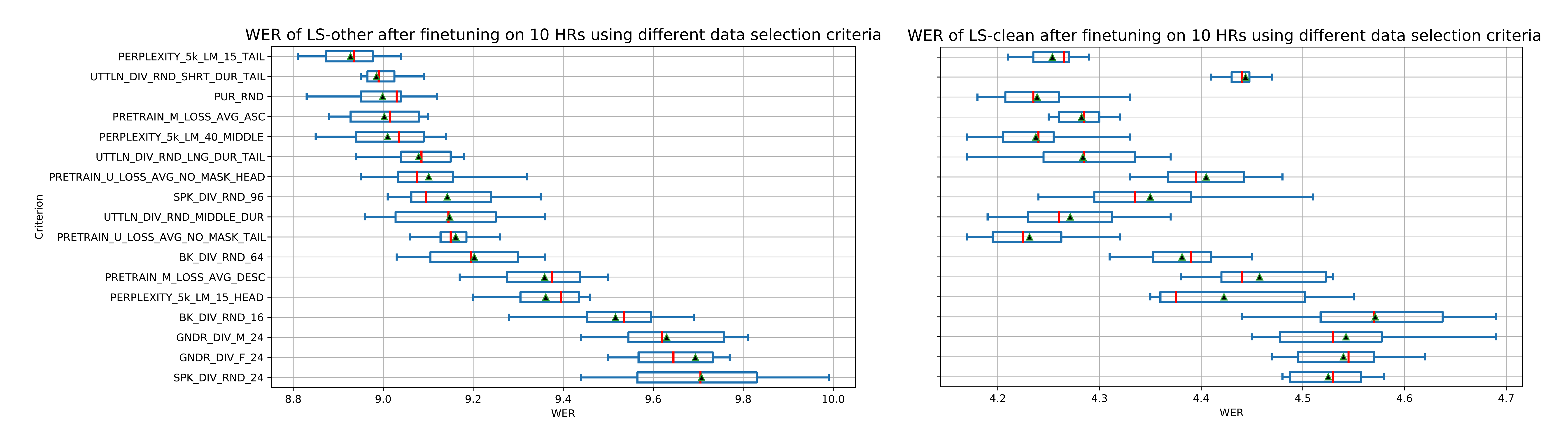}
    \caption{Box plot showing the WER on test-clean and test-other for different data selection criteria. The green triangle represents the mean, while the red line represents the median. Also shown for each criteria are the minimum, maximum, 25th percentile, and 75th percentile.}
    \label{fig:WER_boxplot}
\end{figure*}
\subsection{Training}
 We fine-tune the pre-trained HuBERT base model with target letter labels using each of the selected subsets  described in table \ref{tab:selection_criteria} for 25k updates . Similar to \cite{hsu2021hubert}, we fix the convolutional waveform encoder for the whole training. We freeze the transformer encoder for the first 10k updates and then allow it to train with the rest of the model for the remaining updates. We use adam optimizer with betas set to 0.9 and 0.98 for model optimization. We fine-tune using a two stage learning rate scheduler, where the model ramps up to a peak learning rate of 2e-5 for the first 8k updates and then decays for the remaining updates. We use 2 gpus, and set the batch size to a maximum of 3200000 frames per gpu, and allow padding to the length of the longest utterance per batch. We start validation after 10k updates and keep the best checkpoint on dev-other subset.

\subsection{Results}
For each data selection criterion, we create 8 different random subsets and fine-tune the model on each of these subsets. We evaluate the best checkpoint on dev-other for each fine-tuning subset and then record the mean, minimum,  maximum, median, 25th percentile, and 75th percentile WER on both test-clean and test-other for each group of 8 random subsets. A 4-gram language model trained on Librispeech is used for all the decoding experiments. The box plots in Figure~\ref{fig:WER_boxplot} summarize the results of our experiments. 

For PBPE, we see that for test-other, we score 8.93\% WER on average when selecting from the 15\% of utterances with the highest perplexity score (TAIL). On the other hand we observe that the mean WER degrades to 9.36\% when sampling from the 15\% of the utterances with the lowest perplexity score (HEAD). 
We have a similar observation for test-clean, where the mean WER for TAIL is 4.25\% compared to 4.42\% for HEAD.
For average masked pre-training loss, fine-tuning with the utterances with the smallest loss leads to a mean WER of 9.00\% and 4.28\% on test-other and test-clean respectively. However, using the largest loss utterances in fine-tuning leads to a mean WER of 9.36\% on test-other and 4.46\% on test-clean.
For average unmasked pre-training loss, the mean WER on test-clean is 4.23\% when sampling the fine-tuning subsets from the 15\% utterances with the largest loss (TAIL), while it degrades to 4.41\% when sampling from the 15\% utterances with the smallest loss (HEAD). For test-other the mean WER is 9.1\% for HEAD and 9.16\% for TAIL, which is almost the same.
We observe that the best results obtained from our proposed criteria are almost on par with the pure random selection criterion which scores a mean WER of 9.00\% and 4.24\% on test-other and test-clean respectively.

Moreover, our experiments emphasize the importance of speaker diversity when selecting data for fine-tuning. We see that increasing the number of speakers from 24 to 96 leads to a significant boost in mean WER. It is interesting to see that the mean WER does not vary much regardless of whether the 24 speakers were of the same gender or selected randomly which suggests that the learnt representations from SSL may be somewhat tolerant to gender bias in the fine-tuning data.
However, we do find that topic diversity is crucial as it enriches the vocabulary that is present in the audio. Accordingly, the mean WER significantly improves when increasing the number of different audiobooks sampled for fine-tuning from 16 to 64. Lastly, when sampling from the shortest 15\% of the utterances, the mean WER on test-other significantly improves, but the opposite happens on test-clean. 
In the next section, we dig deeper to analyze these results.



 

\begin{figure}[h!]
    \centering
    \includegraphics[width=\linewidth]{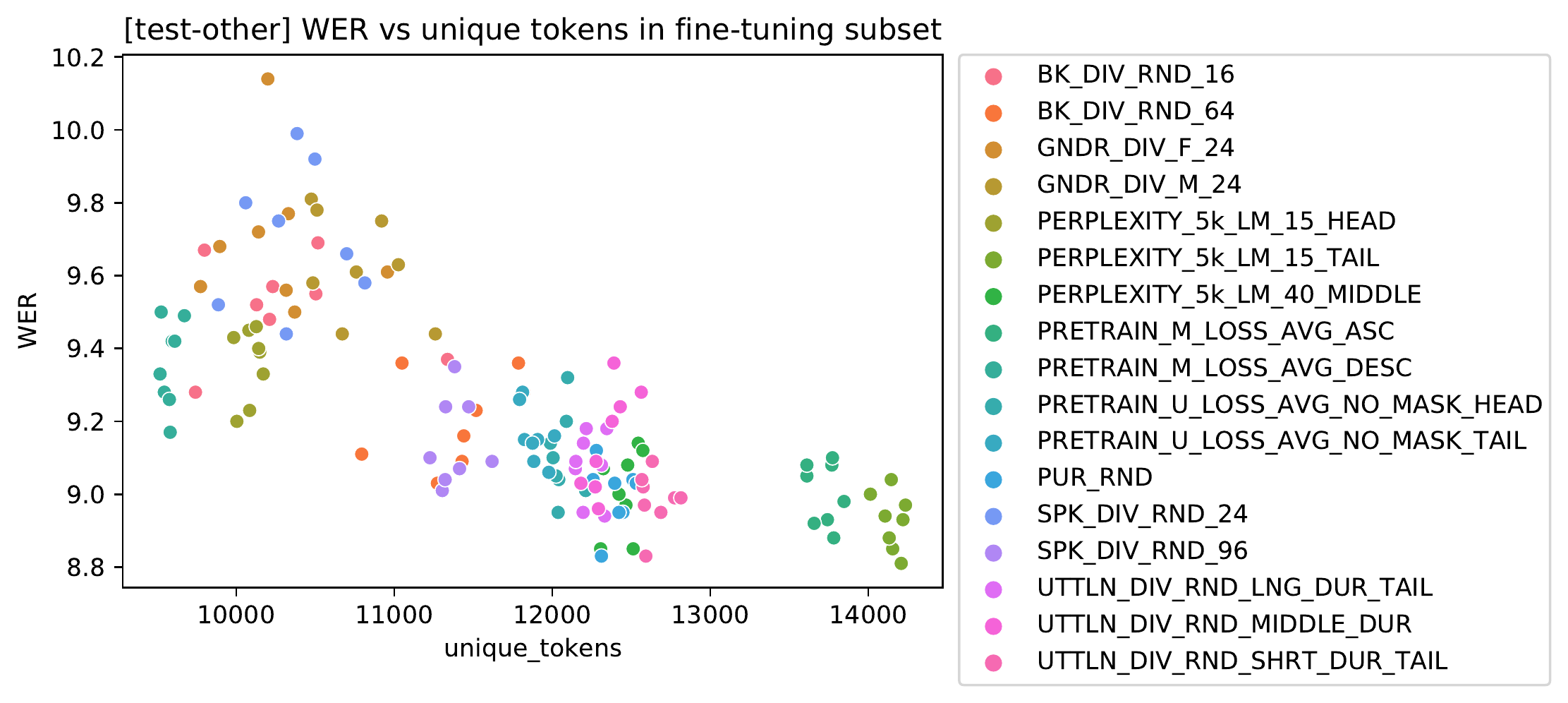}
    \caption{Correlation between WER on test-other and number of unique vocabulary words appearing in the fine-tuning subset}
    \label{fig:unique_tokens_wer}
\end{figure}

\begin{table*}[t!]
    \centering
    \footnotesize
    \begin{tabular}{|l|c|c|c||c|c|c|}
\toprule Fine-tuning subset &WER-other&WERR over &WERR over &WER-clean&WERR over &WERR over \\
  & & libri-light & PUR\_RND &  & libri-light & PUR\_RND\\
\hline
Libri-light&9.61&0.00\%&-6.78\%&4.48&0.00\%&-5.66\%\\
\hline
PPL&8.93&\textbf{7.08\%}&\textbf{0.78\%}&4.25&\textbf{5.05\%}&-0.32\\
\hline
PPL+speaker diversity&8.89&\textbf{7.49\%}&\textbf{1.22\%}&4.06&\textbf{9.38\%}&\textbf{4.25\%}\\
\hline
PPL+book diversity&8.8&\textbf{8.43\%}&\textbf{2.22\%}&4.21&\textbf{6.03\%}&\textbf{0.71\%}\\\bottomrule

    \end{tabular}
    \caption{Results summary for PBPE experiments and fine-tuning with libri-light. In the table, PPL refers to PERPLEXITY\_5k\_LM\_15\_TAIL. The WER results for our criteria are the mean WER computed over 8 runs. WERR over libri-light refers to the word error rate reduction obtained when using each fine-tuning subset relative to fine-tuning with libri-light. WERR over PUR\_RND refers to the word error rate reduction obtained when using each fine-tuning subset relative to fine-tuning with a randomly selected subset.}
    \label{tab:oracle_results}
\end{table*}

\subsection{Analysis}
\begin{figure}
    \centering
    \includegraphics[width=\linewidth]{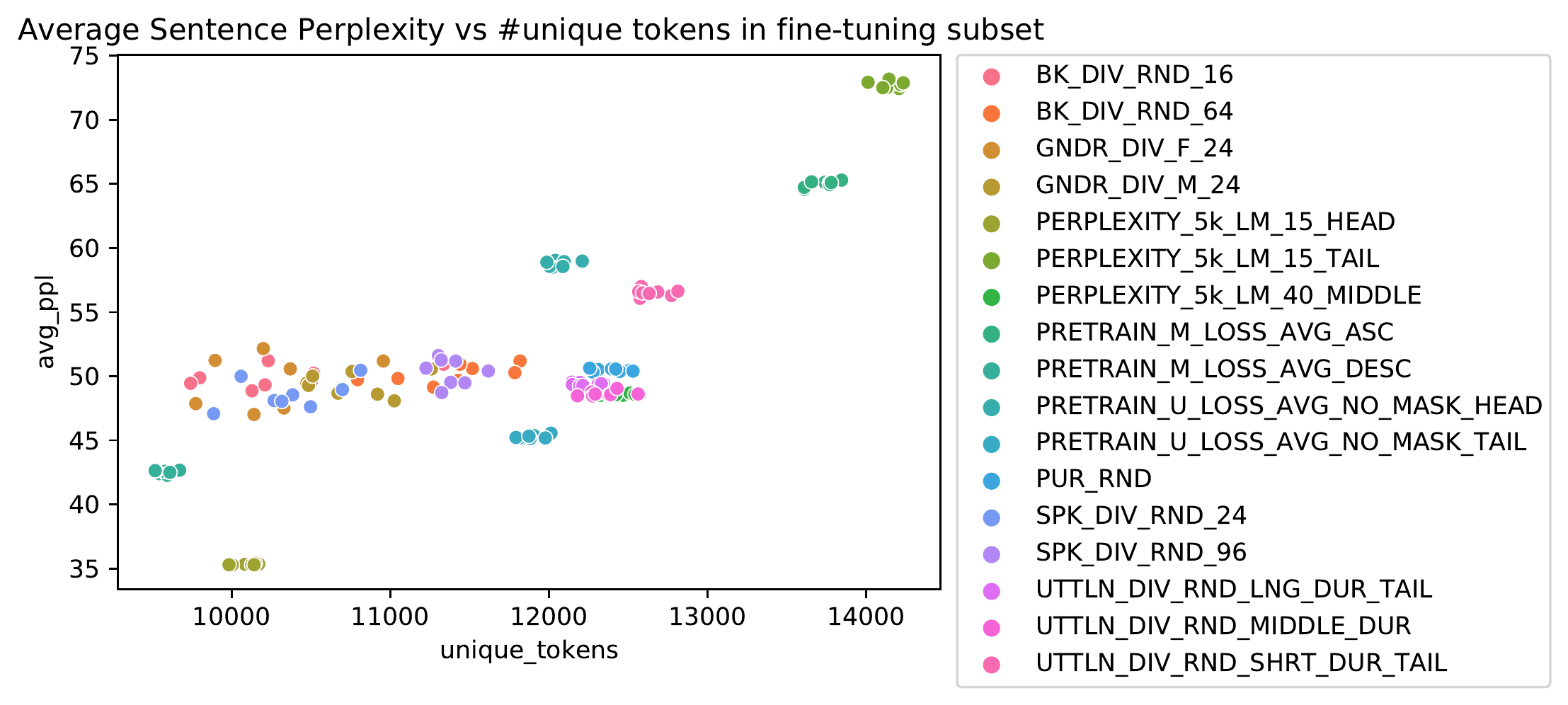}
    \caption{Correlation between the average sentence perplexity computed over the BPE clustered units and the number of unique vocabulary words appearing in the fine-tuning subset}
    \label{fig:ppl_vs_unique_tokens}
\end{figure}

\begin{figure}
    \centering
    \includegraphics[width=.9\linewidth]{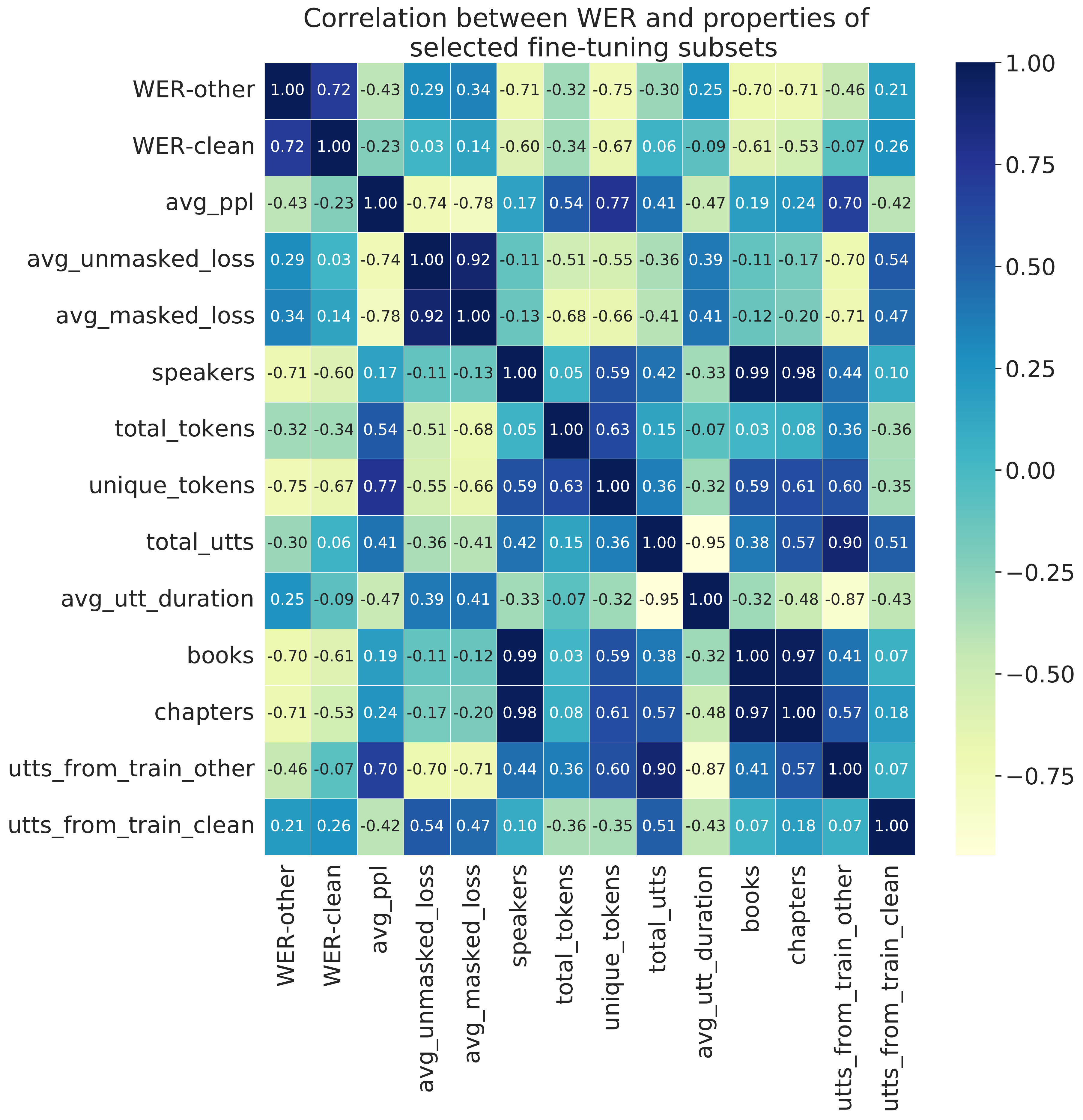}
    \caption{Correlation between the different properties of the selected fine-tuning subsets and the WER on test-other and test-clean}
    \label{fig:correlations}
    \vspace{-.1cm}
\end{figure}

We investigate how the different properties of the selected subsets correlate with each other and with the WER. Figure \ref{fig:correlations} shows some interesting relations between the underlying properties of the selected subsets and the WER observed on both test-clean and test-other. 
We observe a strong negative correlation  between the WER on both test sets and the number of unique vocabulary words in the fine-tuning subset. This correlation is even stronger than the correlation between the WER and the total number of vocabulary words in these subsets. It is obvious that a strong positive correlation exists between the number of unique vocabulary words and the perplexity computed over the BPE clustered units (PBPE).  To compute this correlation, we averaged the perplexity over the total number of utterances in each fine-tuning subset (avg\_ppl). Figure \ref{fig:ppl_vs_unique_tokens} highlights this correlation. This correlation suggests that sampling from the higher perplexity scoring utterances leads to more unique vocabulary words in the selected fine-tuning subset. This is an interesting observation because in our unsupervised selection setup, we have no access to the transcription tokens, however we have access to the HuBERT clustered units that we can use as a proxy for the text. 
Figure \ref{fig:unique_tokens_wer} shows the correlation between the WER on test-other and the number of unique vocabulary words in the fine-tuning subset and how the different selection criteria result in different numbers of unique vocabulary words. Moreover, we observe a strong negative correlation between the number of speakers and the WER and similarly for the number of books. 
Driven by these observations, we conduct two experiments. In the first experiment, we select utterances from all the speakers in the top 15\% highest scoring utterances in terms of perplexity over BPE clustered units. This ensures speaker diversity and maximizes the number of unique vocabulary words, and enables us to beat pure random selection on both test-clean and test-other. In the second experiment, we select utterances from almost every book in the top 15\% highest scoring utterances in terms of perplexity. We arrive at similar results where we are able to beat pure random selection. 
Finally, we compare our results to fine-tuning HuBERT base model with libri-light 10 hour subset \cite{kahn2020libri} and observe that our techniques are scoring better in terms of WER with a large margin. Table \ref{tab:oracle_results} summarizes the results.

In Figure \ref{fig:correlations}, we observe an interesting correlation between the average unmasked loss and the number of utterances selected from train-other or train-clean. As the average unmasked loss decreases, sampling is more biased to the train-other subset, which can account for scoring a lower mean WER on test-other when sampling from the smaller loss utterances compared to the higher loss ones. However, as the average unmasked loss increases, sampling is more biased to train-clean subset, leading to an improved mean WER on test-clean when sampling from the higher loss utterances. In case of the average masked loss, the mean WER on both test-other and test-clean is better when sampling from the lower loss utterances. Despite the correlation directions with the number of utterances from train-clean and train-other being maintained in case of average masked loss, the improved mean WER on test-clean when sampling from lower loss utterances suggests that other factors are involved. Our investigation shows that for average masked loss criterion, we end up with more speakers and more topics when sampling from the lower loss utterances compared to the higher loss utterances. However, the same does not happen for average unmasked loss criterion which can account for the different behaviour. We hypothesize that since all the utterances in our selection pool are included in the pre-training stage of the model, the pre-training loss based criterion will be highly impacted by how frequent each utterance was fed to the model during pre-training as well as the whole pre-training setup causing some biases in the selection process.






\section{Conclusions and future work}
\label{sec:conclusions}
 In this paper, we investigate different criteria for selecting data for fine-tuning a self-supervised speech model to perform ASR. We select in-domain data from a large pool of unlabeled data using unsupervised techniques. We show that HuBERT unit perplexity-based selection allows us to maximize the number of unique vocabulary words in the fine-tuning subset without having access to the transcripts. We also highlight the importance of speaker diversity and topic diversity in achieving optimal performance. Our experiments show that data selection based on pre-training loss can inherit some biases from the pre-training stage depending on how the examples are fed to the model during pre-training. The experiments conducted suggest that pure random selection is a good technique that could be hard to beat as it guarantees topic, vocabulary and speaker diversity as long as the data selection pool is inherently diverse. At the same time, PBPE also show promise in that direction, especially when speaker diversity is guaranteed in the selected subsets.  In the future, we want to investigate perplexity based selection for a target domain when the unlabeled data pool has a diverse set of domains.

\vfill\pagebreak



\bibliographystyle{IEEEbib}
\bibliography{strings,refs}

\begin{thebibliography}{10}

\bibitem{baevski2020wav2vec}
Alexei Baevski, Yuhao Zhou, Abdelrahman Mohamed, and Michael Auli,
\newblock ``wav2vec 2.0: A framework for self-supervised learning of speech
  representations,''
\newblock {\em Advances in Neural Information Processing Systems}, vol. 33, pp.
  12449--12460, 2020.

\bibitem{hsu2021hubert}
Wei-Ning Hsu, Benjamin Bolte, Yao-Hung~Hubert Tsai, Kushal Lakhotia, Ruslan
  Salakhutdinov, and Abdelrahman Mohamed,
\newblock ``Hubert: Self-supervised speech representation learning by masked
  prediction of hidden units,''
\newblock {\em IEEE/ACM Transactions on Audio, Speech, and Language
  Processing}, vol. 29, pp. 3451--3460, 2021.

\bibitem{park2022unsupervised}
Chanho Park, Rehan Ahmad, and Thomas Hain,
\newblock ``Unsupervised data selection for speech recognition with contrastive
  loss ratios,''
\newblock in {\em ICASSP 2022-2022 IEEE International Conference on Acoustics,
  Speech and Signal Processing (ICASSP)}. IEEE, 2022, pp. 8587--8591.

\bibitem{afshan2021sequence}
Amber Afshan, Kshitiz Kumar, and Jian Wu,
\newblock ``Sequence-level confidence classifier for asr utterance accuracy and
  application to acoustic models,''
\newblock in {\em INTERSPEECH 2021}, 08 2021, pp. 4084--4088.

\bibitem{lu2022unsupervised}
Zhiyun Lu, Yongqiang Wang, Yu~Zhang, Wei Han, Zhehuai Chen, and Parisa Haghani,
\newblock ``Unsupervised data selection via discrete speech representation for
  asr,''
\newblock in {\em INTERSPEECH 2022}, 09 2022, pp. 3393--3397.

\bibitem{kawakami2020learning}
Kazuya Kawakami, Luyu Wang, Chris Dyer, Phil Blunsom, and Aaron van~den Oord,
\newblock ``Learning robust and multilingual speech representations,''
\newblock {\em arXiv preprint arXiv:2001.11128}, 2020.

\bibitem{hsu2021robust}
Wei-Ning Hsu, Anuroop Sriram, Alexei Baevski, Tatiana Likhomanenko, Qiantong
  Xu, Vineel Pratap, Jacob Kahn, Ann Lee, Ronan Collobert, Gabriel Synnaeve,
  et~al.,
\newblock ``Robust wav2vec 2.0: Analyzing domain shift in self-supervised
  pre-training,''
\newblock {\em arXiv preprint arXiv:2104.01027}, 2021.

\bibitem{panayotov2015librispeech}
Vassil Panayotov, Guoguo Chen, Daniel Povey, and Sanjeev Khudanpur,
\newblock ``Librispeech: an asr corpus based on public domain audio books,''
\newblock in {\em 2015 IEEE international conference on acoustics, speech and
  signal processing (ICASSP)}. IEEE, 2015, pp. 5206--5210.

\bibitem{kudo2018sentencepiece}
Taku Kudo and John Richardson,
\newblock ``Sentencepiece: A simple and language independent subword tokenizer
  and detokenizer for neural text processing,''
\newblock {\em arXiv preprint arXiv:1808.06226}, 2018.

\bibitem{ott2019fairseq}
Myle Ott, Sergey Edunov, Alexei Baevski, Angela Fan, Sam Gross, Nathan Ng,
  David Grangier, and Michael Auli,
\newblock ``fairseq: A fast, extensible toolkit for sequence modeling,''
\newblock in {\em Proceedings of NAACL-HLT 2019: Demonstrations}, 2019.

\bibitem{kahn2020libri}
Jacob Kahn, Morgane Rivi{\`e}re, Weiyi Zheng, Evgeny Kharitonov, Qiantong Xu,
  Pierre-Emmanuel Mazar{\'e}, Julien Karadayi, Vitaliy Liptchinsky, Ronan
  Collobert, Christian Fuegen, et~al.,
\newblock ``Libri-light: A benchmark for asr with limited or no supervision,''
\newblock in {\em ICASSP 2020-2020 IEEE International Conference on Acoustics,
  Speech and Signal Processing (ICASSP)}. IEEE, 2020, pp. 7669--7673.

\end{thebibliography}

\end{document}